\documentclass[fleqn]{an}
\usepackage{graphicx}
\usepackage{times}
\usepackage{bm}
\Pagespan{43}{50}
\Yearpublication{2011}
\Yearsubmission{2010}
\Month{9}
\Volume{332}
\Issue{1}
\DOI{10.1002/asna.201012345}
\begin{document}
\sloppy

\newcommand{\EQ}{\begin{equation}}
\newcommand{\EE}{\end{equation}}
\newcommand{\EQA}{\begin{eqnarray}}
\newcommand{\EEA}{\end{eqnarray}}
\newcommand{\brac}[1]{\langle #1 \rangle}
\newcommand{\pd}{\partial}
\newcommand{\pdz}{\partial_z}
\newcommand{\DIV}{\vec{\nabla} \cdot }
\newcommand{\CURL}{\vec{\nabla} \times }
\newcommand{\cross}[2]{\boldsymbol{#1} \times \boldsymbol{#2}}
\newcommand{\crossm}[2]{\brac{\boldsymbol{#1}} \times \brac{\boldsymbol{#2}}}
\newcommand{\ve}[1]{\boldsymbol{#1}}
\newcommand{\mean}[1]{\overline{#1}}
\newcommand{\meanv}[1]{\overline{\bm #1}}
\newcommand{\cst}{c_{\rm s}^2}
\newcommand{\nut}{\nu_{\rm t}}
\newcommand{\etat}{\eta_{\rm t}}
\newcommand{\etatz}{\eta_{\rm t0}}
\newcommand{\memf}{\overline{\bm{\mathcal{E}}}}
\newcommand{\memfi}{\overline{\mathcal{E}}_i}
\newcommand{\etaT}{\eta_{\rm T}}
\newcommand{\urms}{u_{\rm rms}}
\newcommand{\Urms}{U_{\rm rms}}
\newcommand{\brms}{B_{\rm rms}}
\newcommand{\Beq}{B_{\rm eq}}
\newcommand{\eu}{\hat{\bm e}}
\newcommand{\xu}{\hat{\bm x}}
\newcommand{\yu}{\hat{\bm y}}
\newcommand{\zu}{\hat{\bm z}}
\newcommand{\Ou}{\hat{\bm \Omega}}
\newcommand{\kef}{k_{\rm f}}
\newcommand{\tauc}{\tau_{\rm c}}
\newcommand{\tauto}{\tau_{\rm to}}
\newcommand{\HP}{H_{\rm P}}
\newcommand{\St}{{\rm St}}
\newcommand{\Sh}{{\rm Sh}}
\newcommand{\Pm}{{\rm Pm}}
\newcommand{\Pe}{{\rm Pe}}
\newcommand{\Rm}{{\rm Rm}}
\newcommand{\Pra}{{\rm Pr}}
\newcommand{\Ra}{{\rm Ra}}
\newcommand{\Ma}{{\rm Ma}}
\newcommand{\Ta}{{\rm Ta}}
\newcommand{\Ro}{{\rm Ro}}
\newcommand{\Rey}{{\rm Re}}
\newcommand{\Co}{{\rm Co}}
\newcommand{\Cost}{\Omega_\star}
\newcommand{\ReLS}{{\rm Re}_{\rm LS}}
\newcommand{\qxx}{Q_{xx}}
\newcommand{\qyy}{Q_{yy}}
\newcommand{\qzz}{Q_{zz}}
\newcommand{\qxy}{Q_{xy}}
\newcommand{\qxz}{Q_{xz}}
\newcommand{\qyz}{Q_{yz}}
\newcommand{\qij}{Q_{ij}}
\newcommand{\Omx}{\Omega_x}
\newcommand{\Omz}{\Omega_z}
\newcommand{\emf}{\bm{\mathcal{E}}}
\newcommand{\emfi}{\mathcal{E}_i}
\newcommand{\nab}{\mbox{\boldmath $\nabla$} {}}
\newcommand{\meanFFFF}{\overline{\mbox{\boldmath ${\cal F}$}}{}}{}
\def\onethird{{\textstyle{1\over3}}}
\def\onehalf{{\textstyle{1\over2}}}
\def\threefourths{{\textstyle{3\over4}}}
\def\ga{\mathrel{\mathchoice {\vcenter{\offinterlineskip\halign{\hfil
$\displaystyle##$\hfil\cr>\cr\sim\cr}}}
{\vcenter{\offinterlineskip\halign{\hfil$\textstyle##$\hfil\cr>\cr\sim\cr}}}
{\vcenter{\offinterlineskip\halign{\hfil$\scriptstyle##$\hfil\cr>\cr\sim\cr}}}
{\vcenter{\offinterlineskip\halign{\hfil$\scriptscriptstyle##$\hfil\cr>\cr\sim\cr}}}}}

\title{Dependence of the large-scale vortex instability on latitude,
  stratification and domain size}

\author{M.J. Mantere\inst{1}\fnmsep\thanks{Corresponding author:
    {maarit.mantere@helsinki.fi}}, P.J. K\"apyl\"a\inst{1,2}, and
  T. Hackman\inst{1,3}}

\titlerunning{Large-scale vortex instability in Cartesian domains}
\authorrunning{M.J. Mantere et al.}

\institute{ 
Department of Physics, PO BOX 64 (Gustaf H\"allstr\"omin katu 2a), 
FI-00014 University of Helsinki, Finland
\and
NORDITA, AlbaNova University Center, Roslagstullsbacken 23, SE-10691 
Stockholm, Sweden
\and Finnish Centre for Astronomy with ESO, University of Turku, 
V\"{a}is\"{a}l\"{a}ntie 20, FI-21500 Piikki\"{o}, Finland}

\received{2011 Aug 31} 
\accepted{2011 Nov 10}
\publonline{2012 Jan 12}


\abstract{%
  In an earlier study, we reported on the excitation of large-scale
  vortices in Cartesian hydrodynamical convection models subject to
  rapid enough rotation. In that study, the conditions of the onset of
  the instability were investigated in terms of the Reynolds (Re) and
  Coriolis (Co) numbers in models located at the stellar North pole. In
  this study, we extend our investigation to varying domain sizes,
  increasing stratification and place the box at different
  latitudes. The effect of the increasing box size is to increase the
  sizes of the generated structures, so that the principal vortex
  always fills roughly half of the computational domain. The
  instability becomes stronger in the sense that the temperature anomaly
  and change in the radial velocity are observed to be enhanced. The
  model with the smallest box size is found to be stable against the
  instability, suggesting that a sufficient scale separation between
  the convective eddies and the scale of the domain is
  required for the instability to work. The instability can be seen
  upto the co-latitude of 30 degrees, above which value the flow
  becomes dominated by other types of mean flows. The instability can
  also be seen in a model with larger stratification. Unlike the
  weakly stratified cases, the temperature anomaly caused by the
  vortex structures is seen to depend on depth.}
\maketitle

% *******************************
\section{Introduction}
% *******************************
\label{sec:intro}

Hydrodynamical Cartesian convection simulations subject to high enough
rotational influence ($\Co\ga3$) and exhibiting large enough Reynolds
number ($\Rey\ga30$) have been reported to generate vortices, the
sizes of which are large compared to the size of the convection cells
(e.g. Chan 2003, 2007; K\"apyl\"a, Mantere \& Hackman, 2011, hereafter
KMH11). In the moderate Coriolis number regime, the vortices are
cyclonic, suppressing the energy transport by convection, and thereby
appearing as cooler than their surroundings. When Coriolis number is
increased even further, anticyclonic vortices are preferred,
enhancing the convective energy transport, making the vortices appear
as regions warmer than their surroundings (KMH11).

In our previous study (KMH11), we proposed that such vortical
structures could be responsible for cool/hot starspots in rapidly
rotating late-type stars possessing outer convection zones. We were
prompted to look into such a possibility due to the decorrelation of
the surface temperature maps, obtained by Doppler-imaging techniques,
from the distribution of surface magnetic fields, derived through
Zeeman-Doppler imaging methods (e.g. Donati \& Collier Cameron 1997;
Donati 1999; Hussain et al.\ 2000; Jeffers et al.\ 2011; Kochukhov et
al.\ 2011).

The resulting temperature anomaly due to the vortices was shown to be
of the order of 5 percent, being somewhat weaker than the temperature
contrasts deduced from observations. The model, however, was very
simple: for example, the density stratification in the radial
direction was only of the order of 23. Due to the low growth rate of
the instability, requiring several thousand turnover times to
saturate, only a very limited parameter range was studied. In this
study, we extend the previous one by investigating a model with a
larger stratification, models with varying domain size, and place the
computational domains at different latitudes.

% *******************************
\section{Model}
% *******************************
\label{sec:model}

Our model is based on that used by K\"apyl\"a et al.\ (2009) and KMH11. 
A rectangular portion of a star is modeled by a box situated at
colatitude $\theta$. The box is divided into three layers: an upper
cooling layer, a convectively unstable layer, and a stable overshoot
layer (see below). We solve the following set of equations for
compressible hydrodynamics:
\begin{equation}
\frac{\mathrm{D} \ln \rho}{\mathrm{D}t} = -\DIV{\bm U},
 \end{equation}
\begin{equation}
 \frac{\mathrm{D} \bm U}{\mathrm{D}t} = -\frac{1}{\rho}{\bm \nabla}p + {\bm g} - 2\bm{\Omega} \times \bm{U} + \frac{1}{\rho} \bm{\nabla} \cdot 2 \nu \rho \mbox{\boldmath ${\sf S}$}, \label{equ:UU}
 \end{equation}
\begin{equation}
 \frac{\mathrm{D} e}{\mathrm{D}t} = - \frac{p}{\rho}\DIV {\bm U} + \frac{1}{\rho} \bm{\nabla} \cdot K \bm{\nabla}T + 2 \nu \mbox{\boldmath ${\sf S}$}^2 - \frac{e\!-\!e_0}{\tau(z)}, \label{equ:ene}
 \end{equation}
where $\mathrm{D}/\mathrm{D}t = \pd/\pd t + \bm{U} \cdot \bm{\nabla}$ 
is the advective time derivative,
$\nu$ is the kinematic viscosity, $K$ is the heat 
conductivity, $\rho$ is the density, $\bm{U}$ is the
velocity, $\bm{g} = -g\hat{\bm{z}}$ is the gravitational acceleration,
and $\bm{\Omega}=\Omega_0(-\sin \theta,0,\cos \theta)$ is the rotation vector.
The fluid obeys an ideal gas law $p=(\gamma-1)\rho e$, where $p$
and $e$ are the pressure and the internal energy, respectively, and
$\gamma = c_{\rm P}/c_{\rm V} = 5/3$ is the ratio of
the specific heats at constant pressure and volume, respectively.
The specific internal energy per unit mass is related to the
temperature via $e=c_{\rm V} T$.
The rate of the strain tensor $\mbox{\boldmath ${\sf S}$}$ is given by
\begin{equation}
{\sf S}_{ij} = \onehalf (U_{i,j}+U_{j,i}) - \onethird \delta_{ij} \DIV \bm{U}.
\end{equation}
The last term of Eq.~(\ref{equ:ene}) describes the cooling at the top
of the domain.  Here $\tau(z)$ is a cooling time which has a profile
smoothly connecting the upper cooling layer and the convectively
unstable layer below, where $\tau\to\infty$.

The positions of the bottom of the box, bottom and top of the
convectively unstable layer, and the top of the box, respectively, are
given by $(z_1, z_2, z_3, z_4) = (-0.85, 0, 1, 1.15)d$, where $d$ is
the depth of the convectively unstable layer. 
In the case of larger stratification (Set~C), the corresponding
vertical positions read $(z_1, z_2, z_3, z_4) = (-0.4, 0, 1, 1.1)d$,
resulting in a vertical extent somewhat smaller than in Sets~A and D.
Initially the stratification is piecewise polytropic with polytropic
indices $(m_1, m_2, m_3) = (3, 1, 1)$, which leads to a convectively
unstable layer above a stable layer at the bottom of the domain.  In a
system set up this way, convection transports roughly 20 per cent of
the total flux.  Due to the presence of the cooling term, a stably
stratified isothermal layer forms at the top. The standard horizontal
extent of the box, $L_{\rm H}\equiv L_x=L_y$, is $4d$; the horizontal
domain size is varied from half of this to double the size in Set~A.
In Set~C with larger stratification, the horizontal extent of the box
is $5d$.
The simulations in Sets~A and C are made at the North pole,
corresponding to $\theta=0\degr$, while in Set~D, the co-latitude is
varied with coarse steps to cover the latitude range down to
$\theta=60\degr$. The simulations were performed with the {\sc Pencil
  Code}\footnote{http://code.google.com/p/pencil-code/}, which is a
high-order finite difference method for solving the compressible
equations of magnetohydrodynamics.

\subsection{Units and non-dimensional parameters}
Non-dimensional quantities are obtained by setting
\begin{eqnarray}
d = g = \rho_0 = c_{\rm P} = 1\;,
\end{eqnarray}
where $\rho_0$ is the initial density at $z_2$. The units of length, time,
velocity, density, and entropy are
\begin{eqnarray}
&& [x] = d\;,\;\; [t] = \sqrt{d/g}\;,\;\; [U]=\sqrt{dg}\;,\;\; \nonumber \\ && [\rho]=\rho_0\;,\;\; [s]=c_{\rm P}.
\end{eqnarray}
We define the Prandtl number and the Rayleigh
number as
\begin{eqnarray}
\Pra=\frac{\nu}{\chi_0}\;,\;\; \Ra=\frac{gd^4}{\nu \chi_0} \bigg(-\frac{1}{c_{\rm P}}\frac{{\rm d}s}{{\rm d}z
} \bigg)_0\;,
\end{eqnarray}
where $\chi_0 = K/(\rho_{\rm m} c_{\rm P})$ is the thermal
diffusivity, and $\rho_{\rm m}$ is the density in the middle of
the unstable layer, $z_{\rm m} = \onehalf(z_3-z_2)$. The entropy 
gradient, measured at $z_{\rm m}$, in the non-convective hydrostatic state,
is given by
\begin{eqnarray}
\bigg(-\frac{1}{c_{\rm P}}\frac{{\rm d}s}{{\rm d}z}\bigg)_0 = \frac{\nabla-\nabla_{\rm ad}}{H_{\rm P}}\;,
\end{eqnarray}
where $\nabla-\nabla_{\rm ad}$ is the superadiabatic temperature
gradient with $\nabla_{\rm ad} = 1-1/\gamma$, $\nabla = (\pd \ln T/\pd
\ln p)_{z_{\rm m}}$, and where $H_{\rm P}$ is the pressure scale
height.  The amount of stratification is determined by the parameter
$\xi_0 =(\gamma-1) e_0/(gd)$, which is the pressure scale height at
the top of the domain normalized by the depth of the unstable layer.
We use $\xi_0 =1/3$ in Sets~A and B, which results in a density
contrast of about 23 across the whole domain, and roughly 9 over the
convectively unstable layer. We make one run with higher
stratification (C1), for which $\xi_0=1/6$, resulting in a density
contrast of roughly 230 over the convectively unstable layer. We
define the Reynolds and P\'eclet numbers via
\begin{eqnarray}
{\rm Re} = \frac{\urms}{\nu \kef}\;,\;\; {\Pe} = \frac{\urms}{\chi_0 \kef} = \Pr\ {\rm Re}\;,
\end{eqnarray}
where $\kef = 2\pi/d$ is adopted as an estimate for the wavenumber of
the energy-carrying eddies, and $\urms=\sqrt{3 u_z^2}$. This
definition of $\urms$ neglects the contributions from the large-scale
vortices that are generated in the rapid rotation regime.  Note that
with our definitions $\Rey$ and $\Pe$ are smaller than the usual ones
by a factor of $2\pi$.  The amount of rotation is quantified by the
Coriolis number, defined as
\begin{eqnarray}
{\rm Co} = \frac{2\Omega_0}{\urms \kef}\;. \label{equ:Co}
\end{eqnarray}
We also quote the value of the Taylor number,
\begin{equation}
\Ta=\left(2\Omega_0 d^2/\nu\right)^2,
\end{equation}
which is related to the Ekman number via ${\rm Ek}=\Ta^{-1/2}$.

\subsection{Boundary conditions}

The horizontal boundaries are periodic for all variables. Stress-free
conditions are used for the velocity
at the vertical boundaries.
\begin{eqnarray}
U_{x,z}=U_{y,z}=U_z=0.
\end{eqnarray}
The temperature
is kept constant on the upper boundary and the temperature
gradient
\begin{eqnarray}
\frac{dT}{dz}=\frac{-g}{c_{\rm V}(\gamma-1)(m+1)},
\end{eqnarray}
is held constant at the lower boundary, yielding a constant heat flux
$F_0=-K \pd T/\pd z$ through the lower boundary.

\begin{table*}
\begin{tabular}{lccccccccccccc}
  Run &grid &$L_{\rm H}$ &$\theta$ &$\Delta \rho$ &$T_{\rm cyc}/\overline{T}$ &$\Rey$ &$\Pe$ &$\Pra$ &$\Ra$ &$\Co$ &$\mbox{Ta}$ &$\tilde{F}_0$ &Cyc.\\ \hline
  D1   & $256^2\times 128$ &$1$ &$60$ &  $9$  &  -  & $47$ & $23$ & $0.48$  & $2.0\cdot10^6$ & $5.3$ & $1.0\cdot10^8$   & $1.7\cdot10^{-5}$ & no \\ % 256x128d1
  D2   & $256^2\times 128$ &$1$ &$45$ &  $9$  &  -  & $43$ & $20$ & $0.48$  & $2.0\cdot10^6$ & $5.9$ & $1.0\cdot10^8$   & $1.7\cdot10^{-5}$ & no \\ % 256x128d2
  D3$^{*}$   & $256^2\times 128$ &$1$ &$30$ &  $9$  &  -  & $47$ & $23$ & $0.48$  & $2.0\cdot10^6$ & $5.4$ & $1.0\cdot10^8$   & $1.7\cdot10^{-5}$ & yes (A+C) \\ % 256x128d3
  D4$^{*}$   & $256^2\times 128$ &$1$ &$15$ &  $9$  & -  & $44$ & $21$ & $0.48$  & $2.0\cdot10^6$ & $5.8$ & $1.0\cdot10^8$   & $1.7\cdot10^{-5}$ & yes (A+C)\\ % 256x128d4
  D5$^{*}$   & $256^2\times 128$ &$1$ &$0$ &  $9$  &  -  & $45$ & $22$ & $0.48$  & $2.0\cdot10^6$ & $5.6$ & $1.0\cdot10^8$   & $1.7\cdot10^{-5}$  & yes (C)\\ \hline % 256x128d5
  A1$^{*}$   & $512^2\times 128$  &$2$ &$0$ &  $9$  &  $(0.099)$  & $42$ & $15$ & $0.36$  & $2.0\cdot10^6$ & $8.1$ & $1.8\cdot10^8$   & $1.7\cdot10^{-5}$ & yes \\ % 512x128a1
             &&&&&&&&&&&&&(2 $\times$ A) \\
  A2   & $256^2\times 128$  &$1$ &$0$ &  $9$  &  $0.045$  & $44$ & $16$ & $0.36$  & $2.0\cdot10^6$ & $7.7$ & $1.7\cdot10^8$   & $1.7\cdot10^{-5}$ & yes (A)\\ % 512x128a1_half 
  A3   & $128^2\times 128$  &$0.5$ &$0$ &  $9$  &  -  & $39$ & $14$ & $0.36$  & $2.0\cdot10^6$ & $8.7$ & $1.8\cdot10^8$   & $1.7\cdot10^{-5}$ & no \\ \hline % 512x128a1_quarter
  C1$^{*}$   & $256^2\times 192$  &$1$ &$0$ &  $233$  & $(0.052)$  & $94$ & $48$ & $0.5$  & $1.3\cdot10^7$ & $4.3$ & $2.6\cdot10^8$   & $3.4\cdot10^{-5}$ & yes (A+C) \\ \hline % cyc256x192a0
\end{tabular}
\caption{Summary of the runs. Stars indicate that 
  the simulation has not been run to a saturated state. The amount of 
  stratification, $\Delta \rho$, is measured over the convectively 
  unstable region. The temperature anomaly, $T_{\rm cyc}/\overline{T}$, 
  where $T_{\rm cyc}$ is the extremum of temperature in a cyclonic region 
  and $\overline{T}$ the mean temperature of certain horizontal layer, 
  is measured at the middle of the convective layer, $z_m$. 
  The dimensionless input heat flux at the lower boundary of the box is 
  given by $\tilde{F}_0=F_0/(\rho c_{\rm s}^3)$, where $c_{\rm s}$ is the 
  adiabatic sound speed and $\rho$ is the density, both measured at the lower 
  boundary of the domain. The last column indicates the 
  presence of cyclonic (C), anti-cyclonic (A), or both types (A+C) of 
  vortices. 
}
\label{tab:runs}
\end{table*}

\section{Results}
\label{sec:results}

In an earlier study, we investigated the excitation of large-scale
vortices in Cartesian domains with weak density stratification,
located at the North pole of a rapidly rotating star (KMH11). In that
study, we investigated the limiting Reynolds and Coriolis numbers
(i.e. inverse Rossby numbers) above which the instability was
excited. For the present study, we have extended our analysis to
models with varying computational domain size (Sect.~\ref{domain}),
place the box at different latitudes of the star
(Sect.~\ref{latitude}), and present a model with higher density stratification
(Sect.~\ref{sec:strat}). The most relevant input parameters for the
runs and some essential output quantities are listed in
Table~\ref{tab:runs}. As is evident from this table, some of the runs
generating vortices have not yet reached a completely saturated state
(these runs are marked with stars), even though they have been
integrated for several thousands of turnover times. In such
cases, quantities such as the temperature anomaly between the vortex
and its surroundings, might still be underestimated, and the listed
numbers in Table~\ref{tab:runs} should be regarded as lower limits.

\subsection{Dependence on the domain size}\label{domain}

\begin{figure*}
\resizebox{\hsize}{!}
{\includegraphics{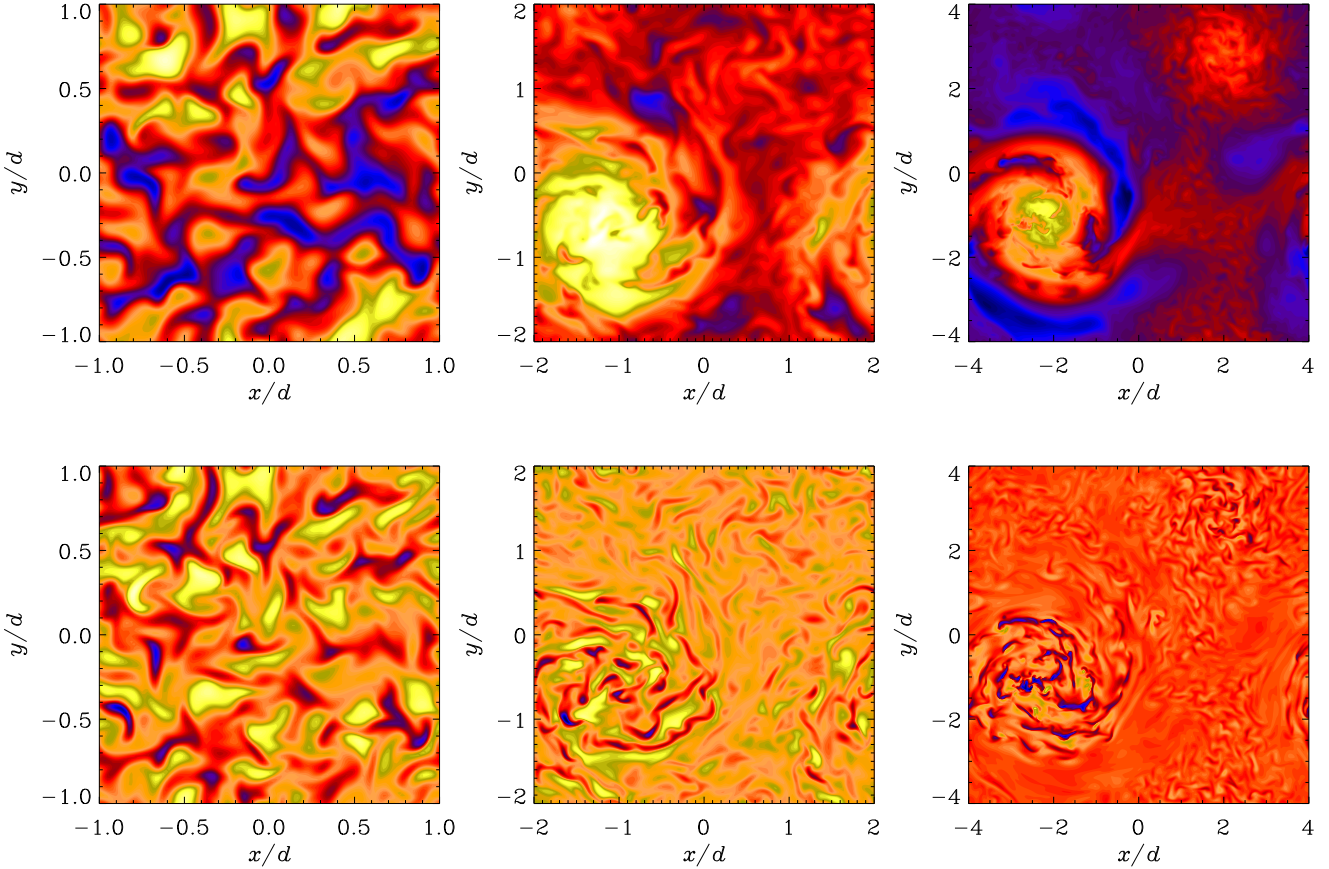}}
\caption{Temperature (upper row) and radial velocity $U_r$ (lower row)
  in the middle of the convectively unstable layer, i.e. at $z_{\rm
    m}$, for Runs~A3, A2 and A1 (from left to right). For Run~A3, the
  slice is taken from the time $t=6400 \tau_{\rm to}$, where
  $\tau_{\rm to}=\left( u_{\rm rms} k_{\rm f} \right)^{-1}$. For
  Runs~A2 and A1, the snapshots are taken at 3400$\tau_{\rm to}$ and
  2700$\tau_{\rm to}$.}
\label{boxsize}
\end{figure*}

In our previous computations with the standard box size of $L_{\rm
  H}$=$4d$ (KMH11), we observed a clear tendency of the sizes of the
vortices to approach the wavenumber $k/k_1$=1, i.e. they tended to
fill in as large a fraction of the horizontal extent as possible. This
prompted us to study the dependence of the instability on the
horizontal extent of our Cartesian box. In Run~A3 of this paper (see
Table~\ref{tab:runs}), both the horizontal extent and resolution of
the domain are halved. Run~A2 is a model with the standard box sixe
(actually otherwise identical to D5 except for the value of the
kinematic viscosity), and in Run~A3 the resolution and horizontal
extent are doubled. In all the runs, the computational domain is
located at the North pole of the star.

In Run~A3, the Reynolds (roughly 39) and Coriolis numbers (roughly
8.7) are clearly above the critical values found in KMH11; still, no
vortices are excited. This is evident from the leftmost panels of
Fig.~\ref{boxsize}, where we show the temperature field (upper panel)
and radial velocity (lower panel) at the middle of the convectively
unstable layer, $z_{\rm m}$. Some large-scale $k/k_1=1$ pattern 
can be detected in the temperature field, that might be
indicative of the early stages of the vortex-instability. This run,
however, was continued up to 6500 turnover times, matching the
timepoint of the slice plotted in Fig.~\ref{boxsize}. It is not
completely ruled out that a very slowly growing vortex instability
mode is present, but in any case its growth rate is strongly reduced
from the standard box runs presented in KMH11.

\begin{figure}
\resizebox{\hsize}{!}{\includegraphics{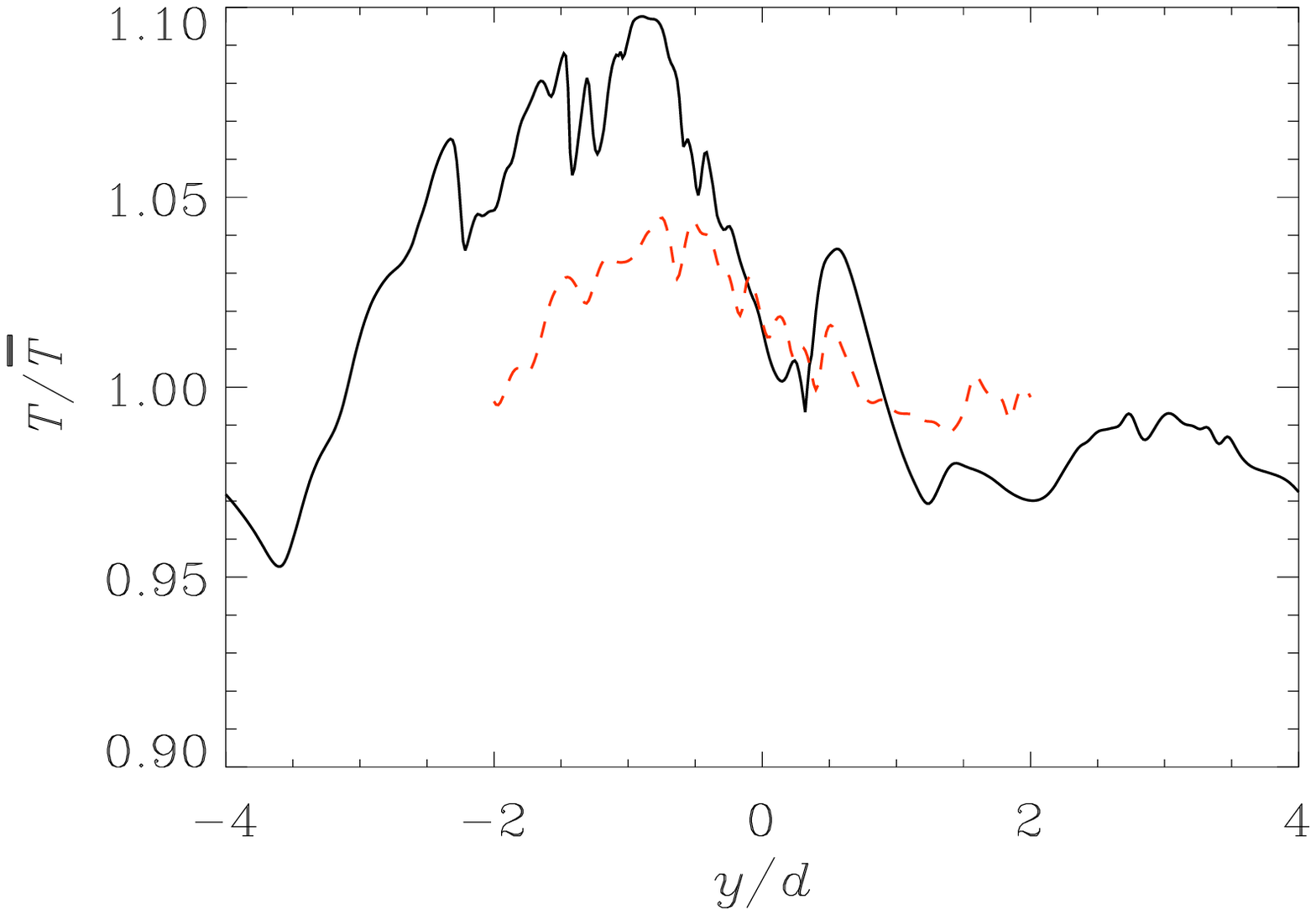}}\\
\resizebox{\hsize}{!}{\includegraphics{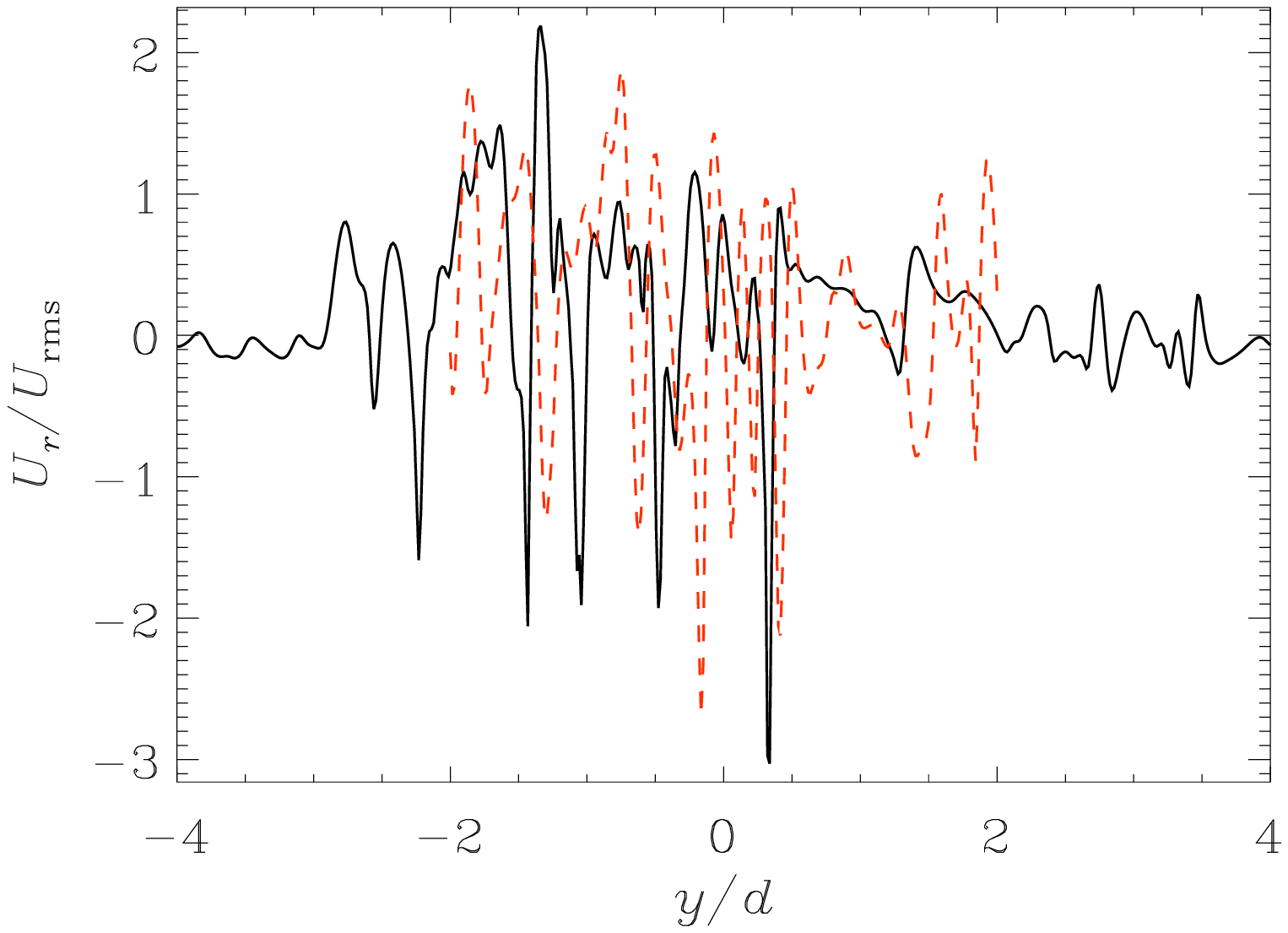}}
\caption{Temperature (upper panel) and vertical velocity $U_r$ (lower
  panel) across the anticyclone for Runs~A1 and A2 with differing box
  size.}
\label{anomalies}
\end{figure}

Run~A2 shows very similar behavior to the earlier calculations
presented in KMH11: the Reynolds number is clearly supercritical to
the instability (roughly 44), and Coriolis number (roughly 7.7) in the
regime where the excitation of an anti-cyclone, i.e. a vortex rotating
in opposite direction to the overall rotation of the domain, was
reported. As shown in the middle panels of Fig.~\ref{boxsize}, a vortex
rotating in the clock-wise direction is seen in the velocity field
(lower panel), the structure being warmer than its surroundings (upper
panel). The temperature anomaly is slightly less than five percent,
very close to the number reported in KMH11. The temperature across the
vortex, normalised to the mean temperature of the horisontal layer, is
plotted in the upper panel of Fig.~\ref{anomalies} with a red, dashed
line. In the region of the anti-cyclone, the vertical velocities
become somewhat enhanced, as evident from the lower panel of the same
figure, where the velocity profile normalised to the rms velocity of
the horisontal layer across the vortex is plotted with a red, dashed
line.

In the model with the largest horizontal extent, Run~A1, exhibiting
very similar Reynolds and Coriolis numbers in comparison to the
standard box Run~A2, two instead of one anticyclones are seen. The
growth rate of the instability, measured from the growth of the
horizontal velocity components, increases by a factor of 2.3 from
Run~A2 to A1. As evident from the rightmost panels of
Fig.~\ref{boxsize}, the two structures are of unequal strength,
possibly suggesting that the system has not yet completely saturated,
although it has been followed for up to nearly 3000 turnover
times. Based on earlier experience with the standard box runs (KMH11),
where multiple vortices were also seen, when followed long enough,
only one vortex very close to the $k/k_1$=1 wavenumber would be the
stable end point for this value of $\Co$.  The temperature and
vertical velocity anomaly across the stronger vortex is plotted in
Fig.~\ref{anomalies} with solid, black lines. As can be seen from this
figure, the heating effect in the middle of the vortex is roughly
twice as large as in the standard box case, suggesting that the
strength of the instability is indeed dependent on the box size. No
such dramatic difference can be seen in the vertical velocity cut
across the vortex (Fig.~\ref{anomalies} lower panel).

\begin{figure}
\resizebox{\hsize}{!}{\includegraphics{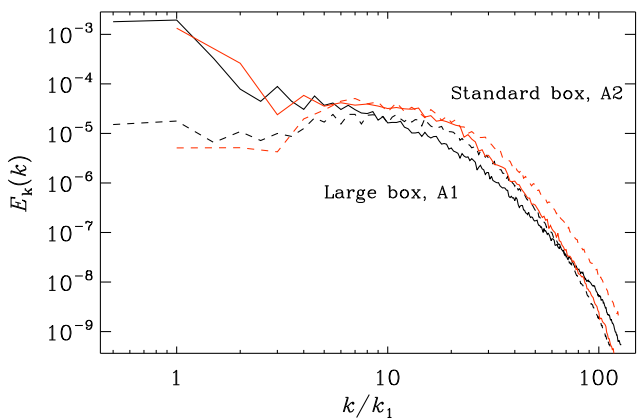}}
\caption{Kinetic energy spectra for the Runs~A1 (black lines) and A2
  (red lines). Spectra taken at early times, when no vortices are yet
  excited, are plotted with dashed linestyle, and spectra from the
  vortex-state with solid linestyle. The wavenumber range of Run~A1 is
  scaled to match the one of Run~A2.}
\label{spectra}
\end{figure}

The size of the stronger vortex in Run~A1, again, is nearly half of
the domain size, i.e. approaching the $k/k_1$=1 mode. In
Fig.~\ref{spectra} we plot the power spectra for the kinetic energy
from Runs~A1 (black lines) and A2 (red lines); the wavenumber scale
for Run~A1 is re-scaled to match the one of Run~A2. In the early
stages, when no vortices are yet excited (dashed linestyles), both the
power spectra consistently peak at intermediate wavenumber of roughly
$k/k_1$=7; this number reflects the size of the turbulent eddies due
to convective motions. Due to the appearance of the vortices, the
energy contained in large scales grows, and dominates the flow in the
nearly saturated stage. Most of the power is seen near the wavenumber
$k/k_1$=1. In Run~A1, the instability would still have 'space' to
advance into even lower wavenumbers i.e. larger scales; followed up
even further, it might still do so.

\subsection{Latitudinal dependence}\label{latitude}

In Set~D, we place the computational domain at different co-latitudes
with a very coarse latitude grid of $\left[0,15,30,45,60\right]$ using
the standard box size. In all the runs in Set~D, the Reynolds and
Coriolis numbers are kept above the critical values found by KMH11. In
Runs~D5, D4 and D3, we still observe the excitation of vortices, but
in the rest, other types of large-scale flows are generated,
suppressing the instability. Such large-scale flows are normally
referred to as banana cells, seen both in Cartesian (Chan 2001;
K\"apyl\"a et al.\ 2004) and spherical geometries (e.g. Brown et al.\
2008; K\"apyl\"a et al.\ 2011a, 2011b).

\begin{figure}
\resizebox{\hsize}{!}
{\includegraphics{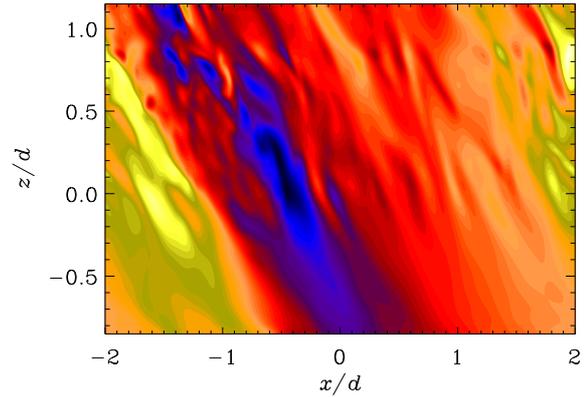}}
\caption{Two-dimensional slice, in the $xz$-plane, of the azimuthal
  velocity $U_y$ from the run D3 at co-latitude 30 degrees. The
  anticyclone generated in the run shows up as a structure spanning
  through the entire vertical extent of the computational domain,
  inclined with the rotation vector.}
\label{theta30}
\end{figure}

The structures extend through the whole convection zone and even
penetrate to the overshoot region, see Fig.~\ref{theta30}, where we
plot a radial-vertical ($xz$) slice of the azimuthal velocity $U_y$ 
from Run~D3 with $\theta$=30$^{\circ}$. The convection and
also the vortex tube, are strongly affected by rotation, which
forces the structures to become inclined with the axis of rotation.

The growth rate of the instability is reduced when the co-latitude is
increased; therefore, it has been extremely difficult to follow the
Runs~D3 and D4 up to saturation. The run at the pole (D5) having the
largest growth rate has been run to a state very near saturation,
when the growth of the horizontal velocity components slows down. In
an earlier state, this run also exhibited both an anti-cyclonic and
cyclonic vortex; in the later stages, however, only the cyclone
persists. The other runs still show both types of vortices, but this
may still change as these runs are relatively further away from the
saturated state in comparison to D5.

\subsection{Dependence on stratification}\label{sec:strat}

We have made an attempt to quantify the effect of increasing
stratification on the vortex instability by running one model where
the stratification is almost 30 times larger than in our standard
cases. Due to this, larger resolution in the radial direction is
needed; the amount of gridpoints has been increased from $nz$=128 to
192, making these computations even more demanding than the rest
included in this study. This run is identified as Run~C1 in
Table~\ref{tab:runs}, and can be observed to have a much higher
Reynolds number (roughly 94) than any other run, but clearly a lower
Coriolis number (roughly 4.3) than the rest of the runs. Nevertheless,
these numbers exceed the critical values found in KMH11, and therefore
the vortex-instability is to be expected, unless the increased
stratification has a significant effect on its excitation conditions.

\begin{figure}
\resizebox{\hsize}{!}
{\includegraphics{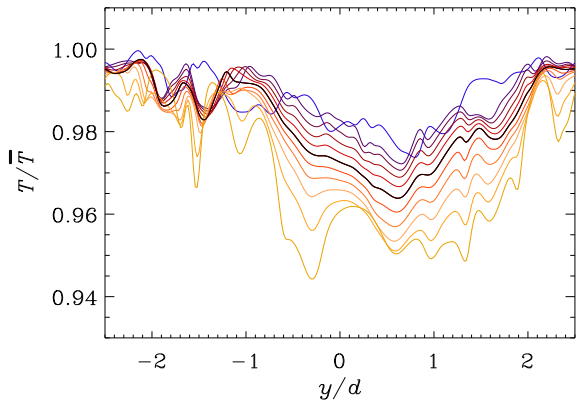}}
\caption{Azimuthal cuts through the computational domain at $x=-1.1$ for
  various depths. The bluer the color, the nearer the bottom of the
  convection zone the cut is taken. Red colors represent cuts near
  $z_{\rm m}$, and the black line a cut taken exactly at $z_{\rm
    m}$. The yellow colors represent cuts taken near the top of the
  convection zone.}
\label{strat}
\end{figure}

Indeed, we observe the instability in Run~C1, generating a stronger
cyclonic and a weaker anti-cyclonic vortex, although the growth rate
of it is reduced in comparison to the other runs due to the lower
Coriolis number. We have been able to calculate this model up to
roughly 2000 turnover times, but it is clear that the system is still
far from saturation. Nevertheless, the temperature anomaly measured at
this relatively early stage is already comparable (maximally close to
5\%) to the standard box calculations very near saturation. The
vortices, again, occur at the very largest scales of the box, and
their size does not vary significantly as a function of depth,
even though the system is more strongly stratified. The temperature
anomaly, on the contrary, varies monotonically through the convection
zone, see Fig.~\ref{strat}, where we plot a temperature cut in the
$y$-direction through a cyclonic, cooler, region located at $x=-1.1$
for various depths. From this figure it is evident that the
temperature contrast is monotonically increasing as function of depth.
In the models with weaker stratification, such an effect is not
clearly visible.

\section{Conclusions}
\label{sec:conclusions}

In an earlier study (KMH11) we reported on the excitation of
large-scale vortices in Cartesian hydrodynamical convection models
subject to rapid enough rotation. In that study, the conditions of the
onset of the instability were investigated in terms of the Reynolds
and Coriolis numbers in models located at the stellar North pole. In
this study, we extend our investigation to varying domain sizes,
increasing stratification and place the box at different latitudes.

The effect of the increasing box size is to increase the sizes of the
generated structures, so that the principal vortex always fills
roughly half of the computational domain. The instability becomes
stronger in the sense that the temperature anomaly and change in the
radial velocity are observed to be enhanced. Also the growth rate,
measured from the time evolution of the horizontal velocity
components, is more than doubled when the boxsize is doubled.  The
model with the smallest box size is found to be stable against the
instability even though the critical Reynolds and Coriolis numbers
found in the earlier study KMH11 are clearly exceeded, suggesting that
a sufficient scale separation between the convective eddies and the
smallest wavenumber of the domain is required for the instability to
work.

The instability can be seen up to a co-latitude of 30 degrees, but at
higher co-latitudes, the flow becomes dominated by large-scale flows
known as banana cells. Such flows have earlier been found from
Cartesian and spherical convection models. The vortices are seen to
align with the rotation vector, forming structures tilted but coherent
through the entire convection zone, extending even to the overshoot
region. The growth rate of the instability is observed to become lower
with increasing co-latitude.

Only very little variance of the temperature contrast across
the vortices can be seen as function of depth when stratification is
small. The instability can also be seen in a model with larger
stratification. Unlike the weakly stratified cases, the temperature
anomaly caused by the vortex sturctures is seen to depend on depth.

\acknowledgements{Computational resources granted by CSC -- IT Center
  for Science, who are financed by the Ministry of Education, and
  financial support from the Academy of Finland grants No.\ 136189,
  140970 (PJK) and 218159, 141017 (MJM), and the `Active Suns'
  research project at University of Helsinki (TH) is acknowledged. The
  authors acknowledge the hospitality of NORDITA during their visits.
}


\begin{thebibliography}{}

%\bibitem[1990]{BTNPS90} Brandenburg, A., Tuominen, I., Nordlund, \AA.,
%  et al.: 1990, A\&A 232, 277
%
%\bibitem[1991]{BMRT91} Brandenburg, A., Moss, D., R\"udiger, G.,
%  Tuominen, I.: 1991, GApFD 61, 179
%
%\bibitem[1992]{BMT92} Brandenburg, A., Moss, D., Tuominen, I.: 1992,
%  A\&A 265, 328
%
%\bibitem[1996]{BJNRST96} Brandenburg, A., Jennings, R.L.,
%  Nordlund, \AA., et al.: 1996, JFM 306, 325
%
%\bibitem[2005]{B05} Brandenburg, A.: 2005, \apj\ 625, 539
%
%\bibitem[2005]{BS05} Brandenburg, A., Subramanian, K.: 2005,
%  PhR 417, 1
%
%\bibitem[2007]{BBBMNT07} Brown, B.P., Browning, M.K., Brun, A.S.,
%  et al.: 2007, in: R.J. Stancliffe, G. Houdek, R.G. Martin, C.A.
%  Tout (eds.), {\it Unsolved Problems in Stellar Physics: A
%  Conference in Honor of Douglas Gough}, AIPC 948, p.~271
%
\bibitem[2008]{BBBMT08} Brown, B.P., Browning, M.K., Brun, A.S.,
  Miesch, M.S., Toomre, J.: 2008, ApJ 689, 1354
%
%\bibitem[2010]{BBMBT10} Brown, B.P., Browning, M.K., Miesch, M.S.,
%  Brun, A.S., Toomre, J.: 2010, ApJ 711, 424 
%
%\bibitem[2006]{BMBT06} Browning, M.K., Miesch, M.S., Brun, A.S.,
%  Toomre, J.: 2006, \apj\ 648, L157
%
%\bibitem[2002]{BT02} Brun, A.S., Toomre, J.: 2002, \apj\ 570, 865
%
%\bibitem[2004]{BMT04} Brun, A.S., Miesch, M.S., Toomre, J.: 2004,
%  \apj\ 614, 1073
%
%\bibitem[2002]{B02} Busse, F.: 2002, PhFl 14, 1301
%
\bibitem[2001]{C01} Chan, K. L. 2001, ApJ, 548, 1102

\bibitem[2003]{C03} Chan, K. L. 2003, in Astronomical Society of the
  Pacific Conference Series, Vol. 293, 3D Stellar Evolution,
  ed. S. Turcotte, S. C. Keller, \& R. M. Cavallo, 168

\bibitem[2007]{C07} Chan, K. L. 2007, AN, 328, 1059

%\bibitem[2003]{DRH03} DeRosa, M.L., Hurlburt, N.E.: 2003, in: S.\
%  Turcotte, S.C.\ Keller, R.M.\ Cavallo (eds.), {\it 3D Stellar
%    Evolution}, ASPC 293, p.\ 229
%
%\bibitem[2006]{DSB06} Dobler, W., Stix, M., Brandenburg, A.: 2006,
%  \apj\ 638, 336

\bibitem[1999]{D99} Donati, J.-F. 1999, MNRAS, 302, 457

\bibitem[1997]{DCC97} Donati, J.-F., \& Collier Cameron, A. 1997,
  MNRAS, 291, 1

%\bibitem[1971]{DR71} Durney, B.R., Roxburgh, I.W.: 1971, SoPh 16,
%  3
%
%\bibitem[1999]{Eea99} Elliott, J.R., Miesch, M.S., Toomre, J.: 1999,
%  \apj 533, 546
%
%\bibitem[2010]{GCS} Ghizaru, M., Charbonneau, P., Smolarkiewicz, P.K.:
%  2010, ApJ 715, L133
%
%\bibitem[1983]{G83} Gilman, P.A.: 1983, ApJS 53, 243
%
%\bibitem[1985]{G85} Glatzmaier, G.A.: 1985, ApJ 291, 300

\bibitem[2000]{Hussain2000} Hussain, G. A. J., Donati, J.-F., Collier
  Cameron, A., \& Barnes, J. R. 2000, MNRAS, 318, 961

\bibitem[2011]{Jeffers2011} Jeffers, S. V., Donati, J.-F., Alecian,
  E., \& Marsden, S. C. 2011, MNRAS, 411, 1301

\bibitem[2004]{KKT04} K\"apyl\"a, P. J., Korpi, M. J., Tuominen,
  I. 2004, A\&A, 422, 793

\bibitem[2009]{Kea09} K\"apyl\"a, P. J., Korpi, M. J., \& Brandenburg,
  A. 2009, ApJ, 697, 1153

\bibitem[2011]{Kea11} K\"apyl\"a, P. J., Mantere, M. J. \& Hackman,
%  T., 2011, ApJ, in press
%PJK: arXiv
  T., 2011, ApJ (in press), arXiv:1106.6029

\bibitem[2011]{KMB11} K\"apyl\"a, P. J., Mantere, M. J., Brandenburg,
  A., 2011b, AN, submitted

\bibitem[2011]{KKGBC11} K\"apyl\"a, P. J., Korpi, M. J., Guerrero, G., Brandenburg, A., Chatterjee, P. 2011a, A\&A, 531, A162 

\bibitem[2011]{Oleg11} Kochukhov, O., Hackman, T., Mantere, M. J.,
  Ilyin, I., Piskunov, N., \& Tuominen, I. 2011, In preparation

%\bibitem[1980]{KR80} Krause, F., R\"adler, K.-H.: 1980,
%  \emph{Mean-field Magnetohydrodynamics and Dynamo Theory}, Pergamon
%  Press, Oxford
%
%\bibitem[2008]{KR08} K\"uker, M., R\"udiger, G.: 2008, J. Phys.:
%  Conf. Ser. 118, 012029
%
%\bibitem[2000]{METCGG00} Miesch, M.S., Elliott, J.R., Toomre, J.,
%  et al.: 2000, \apj\ 532, 593
%
%\bibitem[2006]{MBT06} Miesch, M.S., Brun, A.S.,  Toomre, J.: 2006,
%  \apj\ 641, 618
%
%\bibitem[2008]{MBT08} Miesch, M.S., Brun, A.S.,  Toomre, J.: 2008,
%  \apj\ 673, 557
%
%\bibitem[2009]{MT09} Miesch, M.S., Toomre, J.: 2009, AnRFM, 41, 317
%
%\bibitem[2009]{MTBM09} Mitra, D., Tavakol, R., Brandenburg, A.,
%  Moss, D.: 2009, ApJ 697, 923
%
%\bibitem[2010]{Mea10} Mitra, D., Tavakol, R., K\"apyl\"a, P.J.,
%  Brandenburg, A.: 2010, \apj 719, L1
%
%\bibitem[1978]{M78} Moffatt, H.K.: 1978, \emph{Magnetic Field
%    Generation in Electrically Conducting Fluids}, Cambridge
%  Univ. Press, Cambridge
%
%\bibitem[2005]{R05} Rempel, M.: 2005, \apj 622, 1320
%
%\bibitem[2001]{RC01} Robinson, F.J., Chan, K.L.: 2001, MNRAS 321,
%  723
%
%\bibitem[1989]{R89} R\"udiger, G.: 1989, \emph{Differential Rotation
%    and Stellar Convection: Sun and Solar-type Stars} (Akademie
%  Verlag, Berlin)
%
%\bibitem[2004]{RH04} R\"udiger, G., Hollerbach, R.: 2004,
%  \emph{The Magnetic Universe}, Wiley-VCH, Weinheim
%
%\bibitem[2005]{REKK05} R\"udiger, G., Egorov, P., Kitchatinov,
%  L.L.,  K\"uker, M.: 2005, A\&A 431, 345
%
%\bibitem[2003]{Tea03} Thompson, M.J., Christenseen-Dalsgaard, J.,
%  Miesch, M.S. \& Toomre, J.: 2003, ARA\&A 41, 599

\end{thebibliography}
\end{document}